# Multi-Constituent Simulation of Thrombus Deposition

Wei-Tao Wu[1], Megan A. Jamiolkowski[2,3], William R. Wagner[2,3,4,5], Nadine Aubry[6], Mehrdad Massoudi[7], James F. Antaki[1*]

1. Department of Biomedical Engineering, Carnegie Mellon University, Pittsburgh, PA, 15213, USA
2. McGowan Institute for Regenerative Medicine, Pittsburgh, PA, USA
3. Department of Bioengineering, University of Pittsburgh, Pittsburgh, PA, USA
4. Department of Surgery, University of Pittsburgh, Pittsburgh, PA, USA
5. Department of Chemical Engineering, University of Pittsburgh, Pittsburgh, PA, USA
6. Department of Mechanical Engineering, Northeastern University, Boston, MA, 02115, USA
7. U. S. Department of Energy, National Energy Technology Laboratory (NETL), PA, 15236, USA

Address for correspondence:

James F. Antaki, PhD,

Scott Hall 4N209,

Department of Biomedical Engineering

Carnegie Mellon University,

Pittsburgh, PA 15213.

Phone: 412-268-9857.

Email: antaki@cmu.edu

**Abstract**

In this paper, we present a spatio-temporal mathematical model for simulating the formation and growth of a thrombus. Blood is treated as a multi-constituent mixture comprised of a linear fluid phase and a thrombus (solid) phase. The transport and reactions of 10 chemical and biological species are incorporated using a system of coupled convection-reaction-diffusion (CRD) equations to represent three processes in thrombus formation: initiation, propagation and stabilization. Computational fluid dynamic (CFD) simulations using the libraries of OpenFOAM were performed for two illustrative benchmark problems: *in vivo* thrombus growth in an injured blood vessel and *in vitro* thrombus deposition in micro-channels (1.5mm × 1.6mm × 0.1mm) with small crevices (125μm × 75μm and 125μm × 137μm). For both problems, the simulated thrombus deposition agreed very well with experimental observations, both spatially and temporally. Based on the success with these two benchmark problems, which have very different flow conditions and biological environments, we believe that the current model will provide useful insight into the genesis of thrombosis in blood-wetted devices, and provide a tool for the design of less thrombogenic devices.

**Introduction**

The hemostatic response at the site of vascular injury prevents the loss of blood, but excessive thrombosis may impede or interrupt blood flow to vital organs and tissues. The development of a thrombus in the vasculature is associated with myocardial infarction and stroke, as well as venous thromboembolic disorders.[1,2] Thrombus formation in blood-contacting medical devices is a common cause of failure, and one of the most significant sources of morbidity and mortality.[3,4] For instance, thrombosis in patients receiving ventricular assist devices (VADs) is one of the leading adverse events associated with this therapy, and has raised concerns in the medical community[5–8] and with regulatory bodies such as the FDA[9]. Therefore, there is a critical need for improved understanding of the conditions under which hemostatic pathways may proceed to an excessive and undesirable thrombotic response.

Thrombosis is a complex phenomenon in which a combination of interrelated biochemical and hemodynamic factors result in several cascade reactions causing platelet activation, deposition, aggregation, and stabilization[10–12]. The complexity is accentuated by several feed-forward and feedback mechanisms promoting and inhibiting coagulation reactions. Therefore a comprehensive description of thrombus generation requires a model which can account for interrelated reactions involving platelet activation and aggregation, transport of platelets and chemical species in flow, and the interaction between the formed thrombus and the flow field [10,13–17]. The large number of chemical species and the complexity of cascade reactions make it very difficult to synthesize a comprehensive picture of coagulation dynamics using traditional laboratory approaches [16]. This motivates the pursuit of mathematical models [13,17–23].

For practical objective of predicting thrombosis in blood-wetted medical devices there is inevitably a tradeoff between complexity and utility. An overly simplistic model may fail to account for the essential mechanisms listed above. However, an overly complex model that includes the numerous biochemical species and pathways of coagulation may contain many unidentified parameters, and therefore be intractable or indeterminate to compute. A reasonable compromise was formulated by Sorensen et al. in which a set of convection-reaction-diffusion equations were employed to simulate platelet activation. This model featured a weighted linear combination of agonist concentrations, agonist release and synthesis by activated platelets, platelet-phospholipid-dependent generation of thrombin, and thrombin inhibition by heparin [13,14]. All of the parameters employed in Sorensen's model were available from the experimental literature or by calibration with experimental data. However, this model has several shortcomings limiting its versatility, namely: (1) it limits platelet deposition to an idealized, unchanging surface-fluid boundary and does not account for alterations in the flow field due to thrombus growth, (2) it neglects shear-induced platelet activation, and (3) it does not include thrombus embolization (erosion) due to shear. These limitations motivated the development of a more sophisticated model that could build in these important mechanisms in the thrombotic process.

## Results

*Platelet deposition in blood vessel*

The simulation of thrombus deposition in a blood vessel was modeled by assuming initial laminar flow in a straight, cylindrical blood vessel, shown in Figure 1, that was defined to represent the seminal in-vivo experiments of Begent and Born [24] and Born and Richardson [25]. In these experiments, thrombosis was initiated within a blood vessel of the hamster by injecting ADP at a localized injury site. The diameter and length of the simulated vessel was 0.06mm and 0.5mm, respectively. The mean velocity was $800 \mu m/s$ by prescribing a pressure difference between inlet and outlet, corresponding to a Reynold's number of 0.14. The surface properties were assumed to be those of healthy endothelium, with the exception of the injured injection site, which was assumed to be highly adhesive to platelets. The corresponding reaction rates at the boundary were prescribed as $k_{rpd,b} = 4.0 \times 10^{-5}$m/s and $k_{apd,b} = 4.0 \times 10^{-4}$m/s, and the characteristic embolization shear rate was $\tau_{emb,b} = 1.0 \; dyne \; cm^{-2}$. (For the physical meaning of the terms/parameters, please see the tables in methods section and Supplemental Appendix.) The comparatively large value of $k_{apd,b} = 4.0 \times 10^{-4}$m/s ensured that activated platelets nearby are efficiently captured by the injured site. The platelet deposition rate at the surface, $k_{rpd,b}$ was chosen based on the platelet-collagen deposition rate provided by Sorensen[13,14]. The platelet-platelet aggregation rates, $k_{ra} = 3.0 \times 10^{-6}$ m/s, $k_{aa} = 3.0 \times 10^{-5}$ m/s, were based on Sorensen et al. as well[13,14]. The characteristic embolization shear rate was $\tau_{emb} = 30\ dyne\ cm^{-2}$ based on Goodman et al[21]. The boundary reaction rates for normal, healthy endothelium were set to zero. All remaining parameters are provided in the Table 6. The inlet [RP] and [AP] were prescribed $6 \times 10^{14}$ PLTs/m³ and $6 \times 10^{12}$ PLTs/m³ respectively, based on normal values for rodents.[26,27] The [RP] and [AP] are the unactivated resting platelets and activated platelets in flow respectively.

Figure 2 provides a snapshot of the growing thrombus at t=200s. Comparison of the simulation results with experimental observations of Begent and Born [24] were made on the basis of thrombus growth rate, shape, and size, (see Figure 3-5). Figure 3 illustrates the progression of thrombus growth in the simulation. According to Begent and Born [24], the thrombus grew to a height of 1/3 the lumen diameter in approximately 100-200s, while the same size was obtained after about 150s in our simulation. Figure 5 shows the comparison of the height versus the length of the thrombus observed experimentally and numerically. We found that the model accurately predicted the shape (aspect ratio) of the growing thrombus, which was approximately 2.5:1 (length:height) throughout its growth. The deviation between the experiments and simulations may be attributed to some level of uncertainty in the in vivo experiments. According to Begent and Born[24,25], the data were derived from different experiments where the conditions, such as flow rate and vessel geometry, may have varied slightly. Qualitatively, it was observed that the thrombus grew both upstream and downstream of the injection site, which also agrees with the experimental observations. (See Figure 4.) The upstream growth may be due to the capture of the incoming platelets and the relatively low shear rates; while downstream growth is supplied by platelets activated upstream.

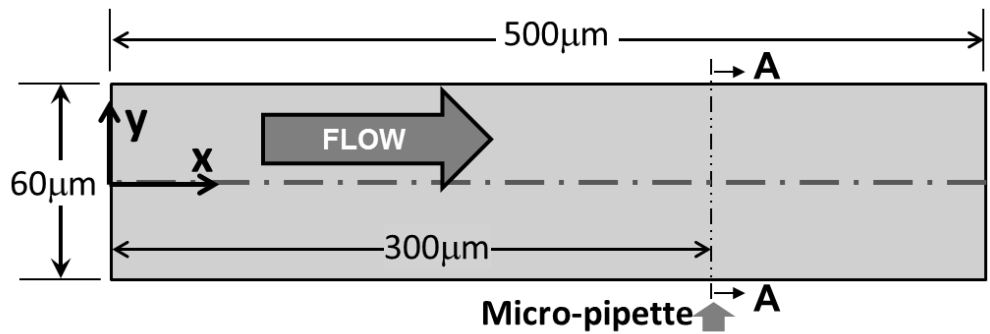

Figure 1 Schematic of the simulated blood vessel of Begent and Born [24] and Born and Richardson [25]. The ADP injection port is located at the bottom vessel wall (diameter = $3\mu m$). The surface properties are assumed to be those of healthy endothelium, with exception of the injection site at which there is an elevated rate of platelet adhesion.

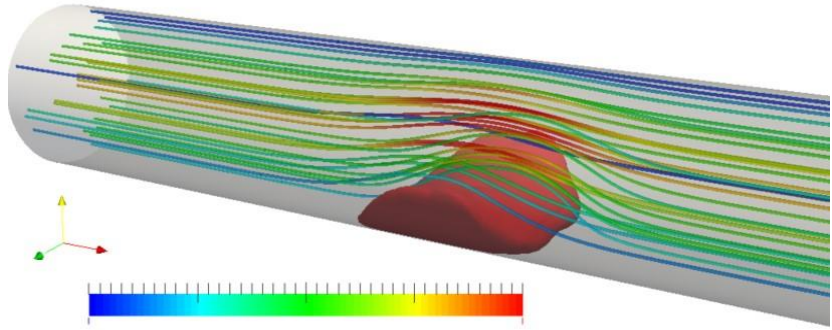

Figure 2 Snapshots of simulated thrombus and streamlines at t=200s with a mean velocity 800$\mu m/s$. The unit of the color scale bar is $\mu m/s$.

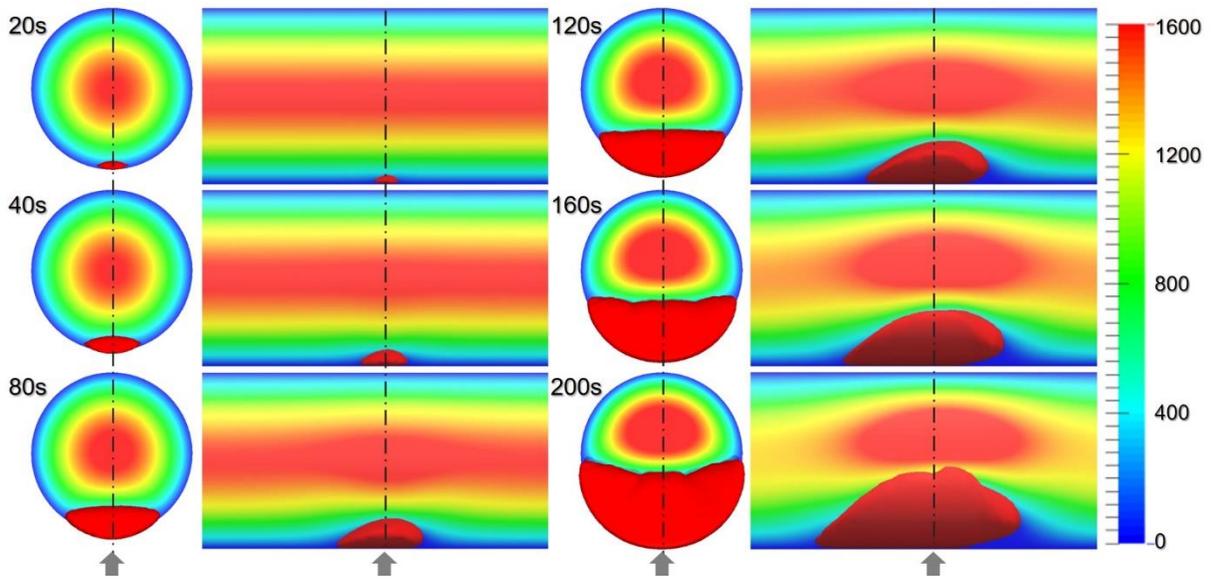

Figure 3 Orthographic views (front and side) of the progression of thrombus growth and velocity field evolution in a circular vessel at a mean velocity 800$\mu m/s$. The side view crosses the plane A-A shown in Figure 1. The arrows indicate the position of the ADP injection port. The unit of the scale bar is $\mu m/s$.

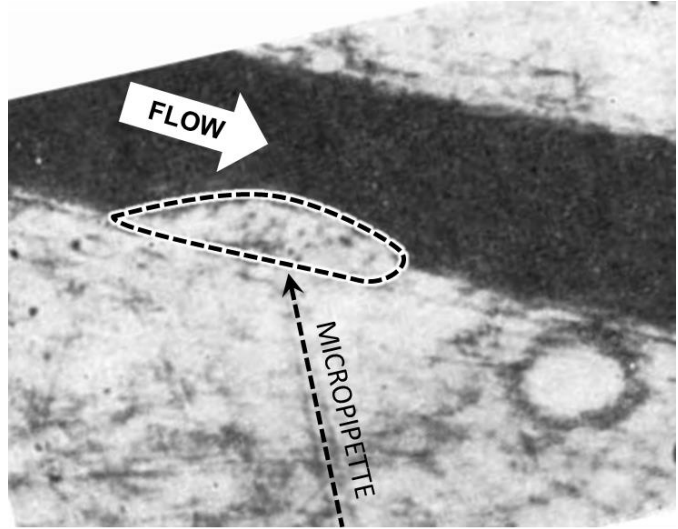

Figure 4 Thrombus in blood vessel observed by Begent and Born [24] after approximately 100-200s of ADP injection through micropipette. The figure is reused by permission.

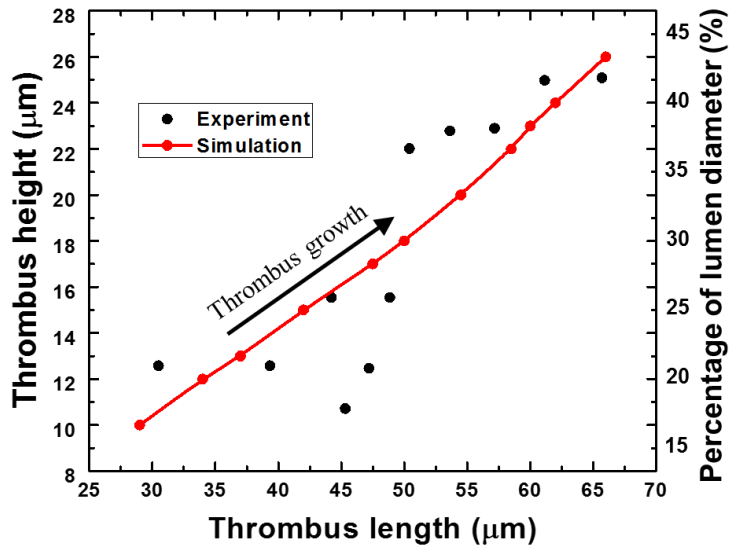

Figure 5 Thrombus height vs length by numerical simulation and experiments at a mean velocity $800 \mu m/s$. Experimental data are from Born et al. [24,25].

Figure 6 shows the streamlines, shear rate, and concentration fields of the 8 remaining species on the radial-axial slice plane. The streamlines indicate that the fluid velocity within the region occupied by the thrombus is zero. Consequently, this produces a stenosis within the vessel. Figure 6(b) indicates a concentration of shear near the surface of the thrombus, which opposes platelet deposition. Figure 6(c) indicates that the concentration of activated platelets in flow is greatest at the leading and trailing edges of the thrombus. Figure 6(e) shows the concentration of the deposited activated platelets where we can find that the concentration in the thrombus core is comparatively higher, which is similar to the observation of Stalker et al. [28]. Figure 6(f) illustrates the diffusion of ADP from the injection site through the thrombus. The concentration of ADP on the surface activates resting platelets which causes the propagation of thrombus growth. Figure 6(g-h) indicates the synthesis of platelet agoinists, TxA$_2$ and thrombin by deposited platelets in the interior of the thrombus. Accordingly Figure 7(i-j) illustrates that prothrombin and anti-thrombin III (ATIII) are consumed in the interior of the thrombus.

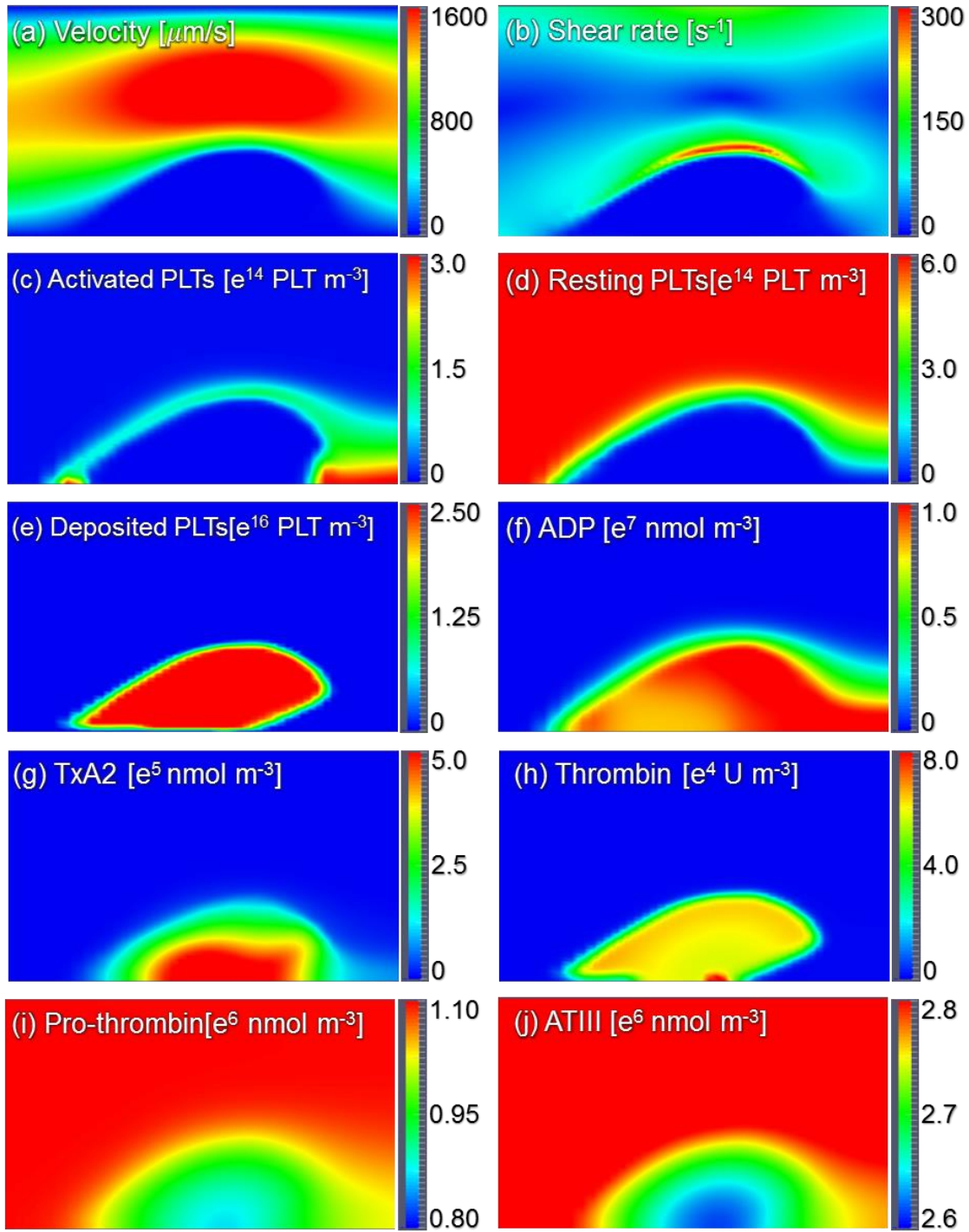

Figure 6 Streamlines, shear rate and concentration fields within the centerline plane at t=200s. It should be noticed that the total activated platelets contain the resting AP shown in figure (c) and deposited AP shown in figure (e).

Additional simulations were performed to study the rate, and time for complete occlusion of the vessel. Table 1 shows that blocking all platelet agonists highly suppresses thrombus growth, and blocking of ADP or Thrombin (TB) increases the occlusion by almost a factor of 2. In contrast, the blocking of $TxA_2$ has a negligible effect on occlusion time. Figure 7 provides the time course of platelet deposition, as measured by the total thrombus volume. We observed that the simulations predict a sigmoidal rate of deposition: initially slow, rapidly increasing until the vessel is completely occluded, at which point the thrombus continues to grow longitudinally at a slower rate. This pattern closely resembles the in-vivo observations of Mangin et al.[28] in the blood vessel of a rat. An additional interesting observation from Figure 7 is that for the case where ADP is blocked, the total thrombus volume at the time of occlusion is greater than the other cases. This implies that for this case, axial thrombus growth occurs more quickly than for the other cases.

Table 1 Time to vessel occlusion by thrombus under varying conditions. For all of the cases there was an injured site as shown in Figure 1, and all unmentioned conditions were the same as for the case studied previously[24,25]. The first case of Begent and Born implies no agonists were blocked, an injured site was present and ADP was injected at the injured site.

| Case | Occlusion time |
| --- | --- |
| ADP injected (same as Begent and Born [24,25].) | 340s |
| No ADP injected. | 380s |
| No ADP injected and with blocking of ADP. | 590s |
| No ADP injected and with blocking of TB. | 620s |
| No ADP injected and with blocking of $TxA_2$. | 380s |
| No ADP injected and with blocking of all agonists. | N/A; (Up to 1200s.) |

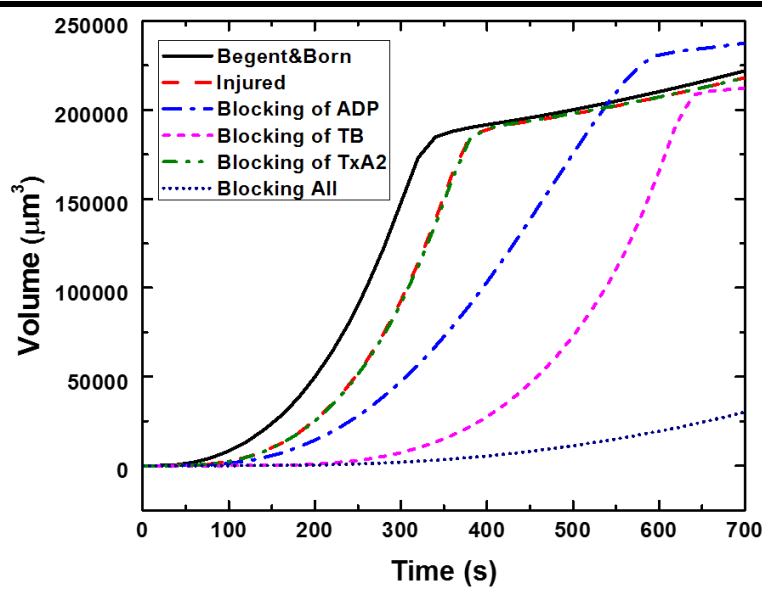

Figure 7 Time course of thrombus formation (volume) to total vessel occlusion with different agonists blocked. The plateaus exhibited by most of the cases correspond to occlusion of the vessel, followed by gradual growth, driven primarily by diffusion.

*Platelet deposition in a micro-crevice*

The second benchmark problem, a rectangular channel with a micro crevice, was motivated by a persistent problem involved in most blood wetted medical devices, namely the seams and joints between component parts that are well known to be predisposed to thrombus deposition. The simulation was formulated to replicate a microfluidic experiment reported previously that utilized human blood [29,30] (See Figure 8). The overall half-height of the channel was 1.5mm, and the depth 0.1mm. The modeled domain only included a portion of the overall 3cm length of the experimental channel. The height of the crevice was 0.125mm and the lengths of the crevice studied here were $L_c = 0.075mm$ and $L_c = 0.137mm$. The inlet velocity was prescribed as 0.0173m/s, corresponding to a Reynolds number of 32.3. The reaction rates and characteristic embolization shear rates were the same as the previous simulation. However the boundary reaction rates and embolization rate needed to be adjusted to account for the different material properties (titanium alloy, Ti6Al4V): $k_{rpdb} = 1.0 \times 10^{-20}$m/s, $k_{apdb} = 1.0 \times 10^{-5}$m/s, $\tau_{embb} = 0.1\ dyne\ cm^{-2}$. The value of $k_{rpdb}$ implies no deposition occurs for resting unactivated platelets, and the values of $k_{apdb}$ and $\tau_{embb}$ were chosen by best fitting the results of the case with $L_c = 0.075mm$. The inlet [RP] and [AP] was prescribed as $2.5 \times 10^{14}$ PLTs/m³ and $1 \times 10^{13}$ PLTs/m³ respectively, corresponding to nominal values for human blood.

Figure 9 shows the deposited platelet field based on the numerical simulation and experiments with the crevice length $L_c = 0.075mm$ at t=300s and 450s. The first column is a three-dimensional rendering of the accumulated thrombus. The second column is the simulated volume fraction of platelets on the near wall, corresponding to the field of view of microscopic observations of fluorescently labelled platelets. [31,32] Comparison of simulation and experiment reveals very good agreement, particularly the thrombus growth at the leading and trailing edges of the crevice. Keeping all the conditions consistent, an additional simulation was performed in channel with a longer crevices ($L_c = 0.137mm$.) Figure 10 shows the corresponding deposited platelets at t= 300 and 450s. Compared to Figure 9, it is clear that lengthening the crevice dramatically reduces the rate of deposition, which is consistent with experimental observations.

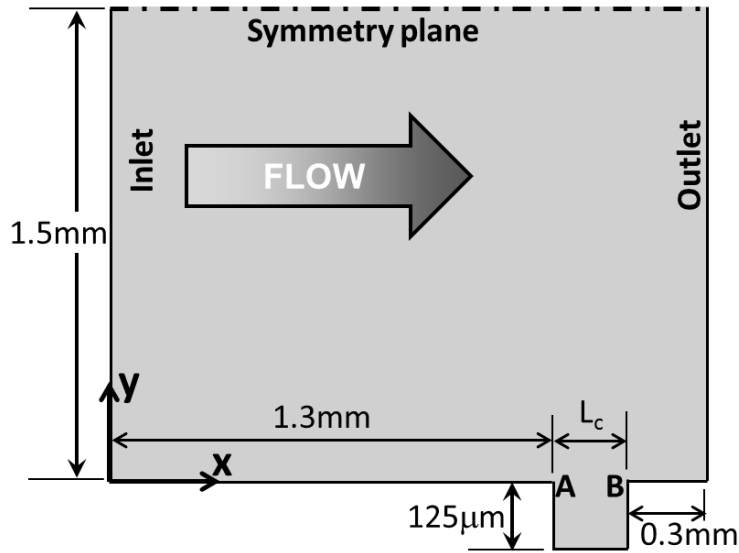

Figure 8 Schematic of the computational domain for rectangular microfluidic channel with a crevice. The overall half-height of the channel was 1.5mm, and the depth is 0.1mm. The domain only included a portion of the overall 3cm length of the experimental channel. The height of the crevice was 0.125mm and lengths, $L_c$, of the crevices were 0.075mm and 0.137mm. The inlet velocity was 0.0173m/s (Re=32.30). A and B indicate upstream and downstream corners of the crevice, respectively.

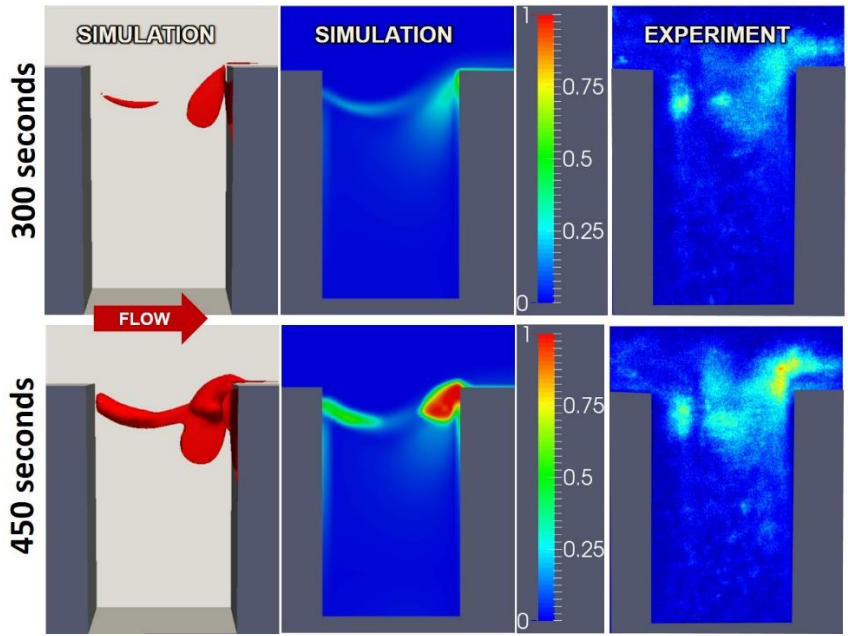

Figure 9 Comparison of thrombus deposition by simulation versus microscopic experiment at t=300 and 450s with crevice length $L_c = 0.075mm$. Left panel is a 3D rendering of the simulated thrombus. Center and right panels indicate volume fraction of platelets in the near-wall region of the channel.

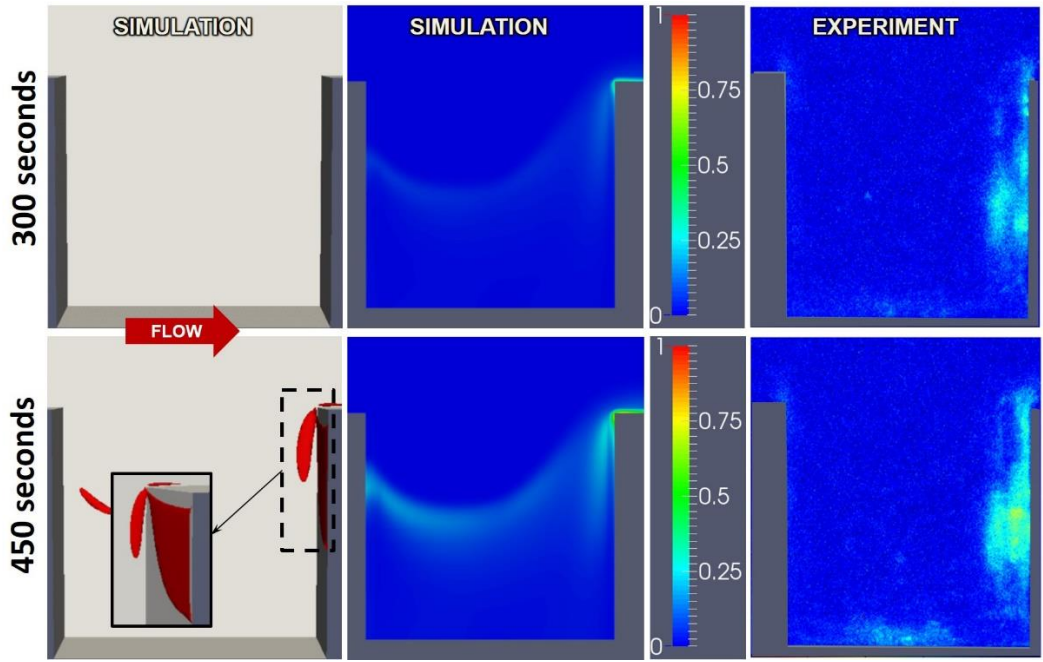

Figure 10 Comparison of thrombus deposition by simulation versus microscopic experiment at t=300 and 450s with crevice length $L_c = 0.137mm$. Left panel is a 3D rendering of the simulated thrombus. Center and right panels indicate volume fraction of platelets in the near-wall region of the channel. The figures are reused by permission. [30]

Figure 11 shows the progression of the simulated thrombus deposition from 300-600s illustrating the growth of both upstream and downstream thrombi with the crevice length of 0.075mm. The downstream corner of the crevice has been indicated as the region where the thrombus grows earliest and fastest, (Figure 11). This may be because at the rear corner, except for the platelets which were activated locally, many platelets are activated without being deposited at the beginning corner. These platelets are transported to the rear corner where they are captured. This makes the growth rate of the thrombus at the rear corner much faster. When the deposited thrombus at the rear corner accumulates to a certain, rather large value, the agonists produced by those deposited platelets become concentrated enough to activate the unactivated surrounding platelets. In this region, the amount of the platelets activated by agonists will be much larger than those activated by the shear stress, which indicates an even faster thrombus deposition at the rear corner. Figure 12 provides the streamlines, shear rate and snapshot of concentration fields of the different species in the model. These figures indicate an accumulation of agonists and activated platelets within the cavity. Once thrombi begin to deposit at the leading and trailing edges, it appears that the secondary vortex within the cavity transports the activated platelets whereupon they accumulate with the deposited platelets in these two regions.

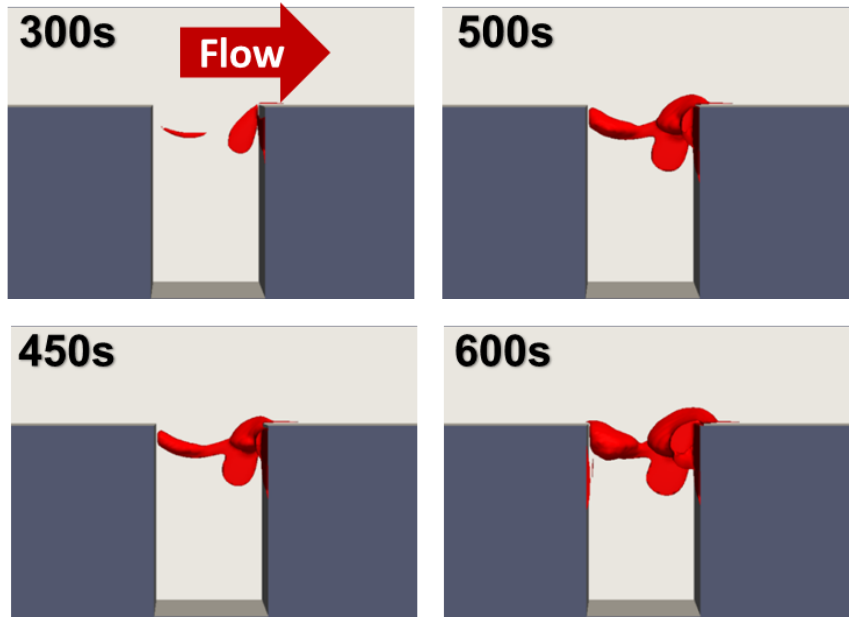

Figure 11 Evolution of thrombus growth in rectangular crevice. The length of the crevice is 0.075mm.

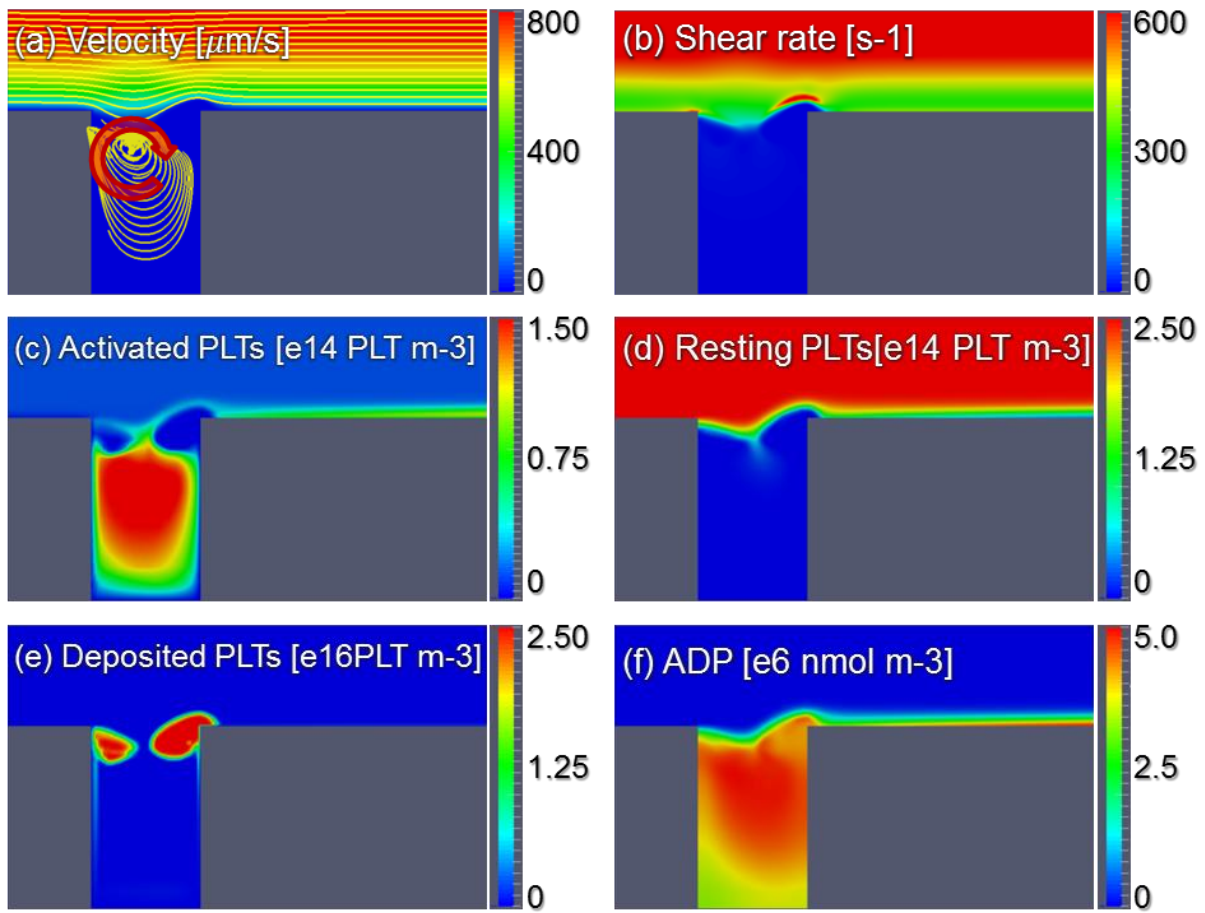

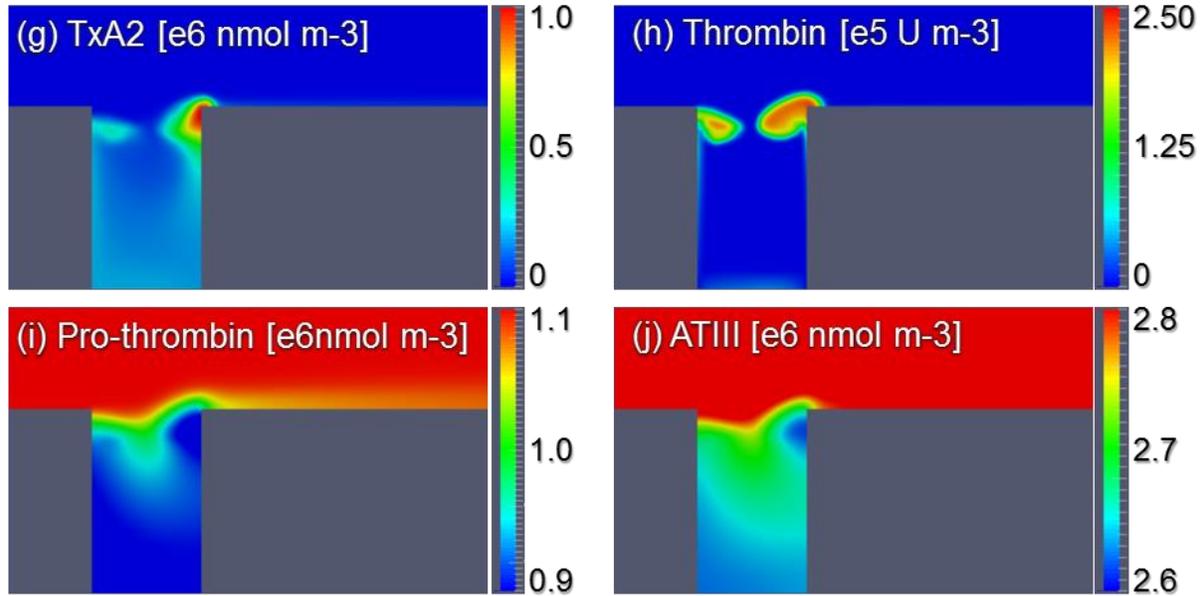

Figure 12 Streamlines, shear rate and concentration fields within the x-y slice plane at distance from the wall $z=10\mu m$, at t=600s. The length of the crevice is 0.075mm.

## Discussion

Mathematical models to simulate blood coagulation and thrombosis have steadily increased in complexity over recent decades.[13,16,20,23,33–35] Paradoxically, parallel efforts have been made to *reduce* the order, and hence complexity, of thrombosis models[36,37]. The appropriate level of complexity depends on the application for which the model is intended, and the questions being asked. For predicting thrombosis in blood wetted devices, the bare minimum that a model should include are the fundamental components of Virchow's triad: (1) the properties of blood, including several states for platelets, biochemical agonists with associated positive feedback mechanisms, and agonist inhibition; (2) the character of flow which includes shear-mediated platelet activation, species convection, and thrombus-fluid interaction; and (3) surface chemistry in terms of surface reactivity.

Models commonly used today also encompass a wide range of physical scales and degree of granularity, each with advantages and disadvantages. In general, micro-scale models are more appropriate for studying phenomena that occur within and between individual cells.[19,20] Such models typically involve many unknown parameters. For this reason, phenomenological models are more practical, yet they are limited by the assumptions necessary to account for un-modeled behavior at the micro-scale. The thrombosis model presented here adopts a continuum approach that seeks to achieve a compromise in complexity and scale. The advantage of this approach is scalability, which enables simulation of arbitrarily shaped, complicated geometries that are encountered with the blood flow pathways of many current medical devices. To account for the most essential cell- and molecular scale mechanisms, this model employs a set of convection-diffusion-reaction equations that represent platelet activation by both chemical agonists and fluid forces; platelet deposition; thrombus propagation; thrombus embolization; thrombus stabilization; and inhibition. This model builds upon many of the principles incorporated in the earlier work of Sorensen et al. [13,14], and inspired by the pioneering work of Fogelson [38]. Although the Sorensen model demonstrated good results in simulating platelet deposition in simple parallel-plate flow [13,14], it lacked several features that limited its versatility. Accordingly, this revised model added shear-induced platelet activation, thrombus embolization due to shear, and fluid-solid interaction with the growing thrombus. The resulting number of species in the transport model was increased from 7 to 10. Similar to Sorensen, the current model derived most parameters from literature. The only parameters that were determined by experimental fitting were the three surface-specific coefficients, namely $k_{rpd,b}$, $k_{apd,b}$ and $\tau_{emb,b}$. If the model were to be used with different materials, such as polycarbonate or stainless steel, these three terms would need to be modified. Likewise, some parameters would need to be adjusted for viscosity, animal species, and various pathologies.

Although this model performed well with the two representative problems studied, it would still benefit from additional validation in different scenarios, such as vascular stenosis [39,40], sudden expansions [32], and other architectures with a broader range of flow conditions and blood properties [19,41,42]. It might be found that some assumptions and simplifications will need to be revisited. For example, we have modeled the blood as a single phase fluid, ignoring the spatial distribution of RBCs. This was done for the sake of numerical stability and in consideration of more complex future applications of the model. It is widely known that platelets tend to concentrate near the vessel wall, due to the interaction with RBCs, which can influence the rate of thrombosis[43–45]. The opposite is also possible: some experimental studies have also found that collisions of RBCs with thrombi reduced their mass[46]. These phenomena may have been compensated to some extent by choosing an appropriate deposition rates in the current work. Furthermore, the current model excluded the influence of von Willebrand factor, which is known to be relevant in platelet deposition in high-shear field devices such as blood pumps.[47–50] Also, the thrombus stabilization rate was assumed to be negligible in the current simulations. Consequently, thrombus growth was limited as deposition reached equilibrium with clearance by shear. However, for problems where deposition occurs over a long period of time, stabilization may play an essential role. Capturing this phenomenon is a challenge, since the process involves several chemical and biological species, such as fibrinogen/fibrin and factor XIII, which would increase the number of parameters. [48,51,52] Such an effort is a focal point for our ongoing work. The reader interested in fibrin formation and thrombus stabilization is referred to recent experimental reports by Colace, Neeves, and Savage [53–55] and the theoretical model of Anand et al. [18].

In its current embodiment, the thrombosis model presented here was able to replicate the patterns of platelet deposition in a micro-crevice observed experimentally. Both experiment and simulation revealed a counter-intuitive growth of thrombus from *both* edges (corners) of the crevice. The simulation however revealed the underlying processes by which an initial platelet deposition serves as a nidus for continued thrombus growth through the activation of resting platelets, generation of thrombin, and a temporal influence on the regional fluid dynamics. These results provide evidence that this model may find application in examining the influence of various hemodynamic and biochemical factors on the undesirable phenomenon of thrombotic deposition in medical devices.

**Conclusions**

We have presented a mathematical model for simulating the initiation and propagation of thrombus formation on biomaterial surfaces. The model demonstrated excellent agreement with two illustrative benchmark problems: *in vivo* thrombus growth in an injured blood vessel and an *in vitro* thrombus deposition in micro-channels with small crevices. Because the model includes several interrelated biochemical and hemodynamic mechanisms, it was also able to simulate the influence of downregulating (blocking) certain biochemical agonists, such as ADP, thrombin, and $TxA_2$. Using the model, we were able to predict the influence of a small crevice in the blood flow path, similar to those frequently found at the seams and joints between component parts of a blood-contacting medical device. By repeated simulations with a range of hemodynamic, biochemical, and geometric conditions, it is hoped that the model will prove useful as part of a broader design process seeking to create more blood biocompatible medical devices.

**Methods: Mathematical Model**

We introduce a mathematical model of thrombosis consisting of equations of motion that determine the pressure and velocity fields and a set of coupled convection-diffusion-reaction (CDR) equations that govern the transport and inter-conversion of chemical and biological species – both within the thrombus and in the free stream.

*Equations of motion*

Blood is treated as a multi-constituent mixture comprised of (1) a fluid phase which is modeled as a linear fluid and (2) a thrombus phase. The fluid phase comprises red blood cells (RBCs) suspended in plasma, and is governed by the following equations of conservation of mass and linear momentum:

$$\frac{\partial \rho_f}{\partial t} + div(\rho_f \boldsymbol{v}_f) = 0 \qquad (1)$$

$$\rho_f \frac{D\boldsymbol{v}_f}{Dt} = div(\boldsymbol{T}_f) + \rho_f \boldsymbol{b}_f - C_2 f(\phi)(\boldsymbol{v}_f - \boldsymbol{v}_T) \tag{2}$$

where $\boldsymbol{T}_f$ is stress tensor of the fluid, represented by:

$$\boldsymbol{T}_f = [-p(1-\phi)]\boldsymbol{I} + 2\mu_f(1-\phi)\boldsymbol{D}_f \tag{3}$$

where $p$ is the pressure, $\mu_f$ is the asymptotic dynamic viscosity (3.5cP). A scalar field $\phi$ is introduced to represent the volume fraction of deposited platelets (thrombus). The density of the fluid phase is defined in terms of the volume fraction according to:

$$\rho_f = (1-\phi)\rho_{f0} \tag{4}$$

where $\rho_{f0}$ is the density of the fluid phase (= 1060 kg/m³), $\boldsymbol{b}_f$ is the body force, $\boldsymbol{v}_f$ and $\boldsymbol{v}_T$ are the velocity of the fluid and thrombus phases, respectively. Therefore the term $C_2 f(\phi)(\boldsymbol{v}_f - \boldsymbol{v}_T)$ is the resistance force on the fluid phase from the thrombus, where the coefficient $C_2 = 2 \times 10^9 kg/(m^3 s)$ is computed by assuming deposited platelets behave like densely compact particles 2.78um in diameter, described by Johnson et al. [56] and Wu et al.[57,58], and $f(\phi) = \phi(1 + 6.5\phi)$ is the hindrance function.

*Convection-diffusion-reaction equations*

The current model includes ten (10) chemical and biological species, summarized in Tables 1 and 2, and illustrated in Figure 1. These include five categories (states) of platelets (1) **RP**: resting platelets (in the flow field); (2) **AP**: activated platelets (in the flow field, and more reactive); (3) **RP_d**: deposited (trapped) resting platelets, (4) **AP_d**: deposited active platelets, and (5) **AP_s**: deposited and stabilized platelets. An additional five biochemical species include (1) **a_pr**: platelet-released agonists (ADP), (2) **a_ps**: platelet-synthesized agonist (thromboxane A_2), which can be degraded via first-order reactions; (3) **PT:** prothrombin; (4) **TB:** thrombin, synthesized from prothrombin on the activated platelet phospholipid membrane; and (5) **AT:** anti-thrombin III, which inhibits thrombin and whose action is catalyzed by heparin via the kinetic model of Griffith. In the current model, value of most parameters are available from literatures.

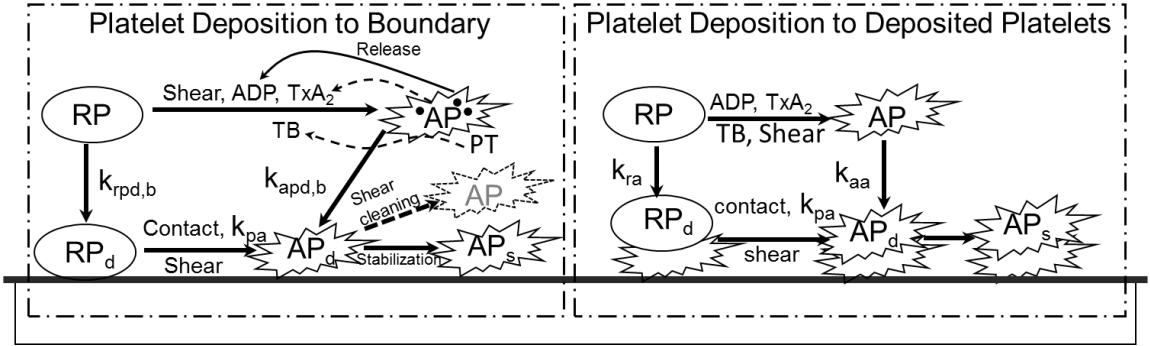

Figure 13 Schematic depiction of the thrombosis model, comprised of platelet deposition, aggregation, and stabilization. RP: resting platelet, AP: activated platelet, $RP_d$ and $AP_d$: deposited resting and active platelets, $AP_s$: stabilized deposited active platelets. Agonists that cause activation (RP to AP) are adenosine diphosphate, ADP; thromboxane $A_2$, $TxA_2$; shear, and thrombin, TB – which is synthesized from prothrombin (PT). The suffix b refers to the reaction with the boundary (surface.) The constants $k_{pa}$, $k_{ra}$, $k_{aa}$, $k_{rpd,b}$, $k_{apd,b}$, refer to the reaction rates for inter-conversion of the associated platelet states.

Figure 1 depicts the fundamental mechanisms that comprise the thrombosis model, which include:

1. *Platelet Activation*: RP can be converted to be AP through exposure to shear[49] and critical levels of biochemical agonists, including ADP[59–61], Thrombin[62,63] and $TxA_2$[64,65]. It is assumed that this conversion results in the release of dense granules (ADP) and causes the platelets to become adherent to biomaterial surfaces and to other platelets.

2. *Platelet Deposition*: Both RP and AP in the free stream are able to deposit to a surface, such as a channel wall, whereupon they are designated $RP_d$ and $AP_d$, respectively. This process is mathematically modeled by a reaction flux at the boundary.

3. *Thrombus Propagation*: When the volume fraction of the deposited platelets ($RP_d$ and $AP_d$) is greater than a specified value, the deposition will propagate downstream. Additional details are provided in Supplemental Appendix.

4. *Thrombus Dissolution or Erosion*: Shear stress by the fluid phase is able to clear deposited platelets, thereby simulating the process known as surface "washing." Mathematically, it was modeled by thrombus clearance term which is related to shear stress.

5. *Thrombus Stabilization*: Newly deposited platelets are converted by a constant rate to be as a solid clot which cannot be removed by hydrodynamic force.

6. *Thrombus Inhibition*: Anti-thrombin III in the free stream and within the thrombus may bind to thrombin, thereby neutralizing the effect of thrombin in activating additional platelets.

7. *Thrombus-fluid interaction*: The permeability of thrombus is inversely related to the volume fraction of deposited platelets. As represented in Equation (2) above, this is computed as a resistance force to fluid flow, analogous to the behavior of packed granular slurries.

The transport of the above species in the flow field is described by a corresponding set of CDR equations [13,14] of the form:

$$\frac{\partial [C_i]}{\partial t} + div(\mathbf{v}_f \cdot [C_i]) = div(D_i \cdot \nabla [C_i]) + S_i \qquad (5)$$

where $[C_i]$ is the concentration of species $i$; $D_i$ refers to the diffusivity of species $i$ in blood; and $S_i$ is a reaction source term for species $i$. Deposited platelets (RP$_d$, AP$_d$, and AP$_s$) are governed by corresponding rate equations:

$$\frac{\partial [C_i]}{\partial t} = S_i \qquad (6)$$

Table 2 and Table 3 lists the appropriate form for the source terms $S_i$, along with the abbreviations and the Table 5 shows the units, diffusion coefficients and normal level of concentrations for $[C_i]$.

Table 2. Source terms associated with platelets. See Figure 1 for definition of reaction rates, k.

| Species | $[C_i]$ abbreviation | $S_i$ form |
|---|---|---|
| Unactivated Resting PLTs | [RP] | $-k_{apa}[RP] - k_{spa}[RP] - k_{rpd}[RP]$ |
| Activated PLTs | [AP] | $k_{apa}[RP] + k_{spa}[RP] - k_{apd}[AP]$ |
| Deposited Resting PLTs | [RP$_d$] | $(1-\theta)k_{rpd}[RP] - k_{apa}[RP_d] - k_{spa}[RP_d] - f_{emb}[RP_d]$ |
| Deposited Activated PLTs | [AP$_d$] | $\theta k_{rpd}[RP] + k_{apd}[AP] + k_{apa}[RP_d] + k_{spa}[RP_d] - (f_{emb} + f_{stb})[AP_d]$ |
| Deposited and stabilized PLTs | [AP$_s$] | $f_{stb}[AP_d]$ |

Table 3. Source terms for chemical species of the model.

| Species | $[C_i]$ abbreviation | $S_i$ form |
|---|---|---|
| PLT-released agonists (ADP) | [a$_{pr}$] | $\lambda_j \big( k_{apa}[RP] + k_{spa}[RP] + k_{apa}[RP_d] + k_{spa}[RP_d] + \theta k_{rpd}[RP] \big) - k_{1,j}[a_{pr}]$ |
| PLT-synthesized agonists (TxA$_2$) | [a$_{ps}$] | $s_{pj}([AP] + [AP_d]) - k_{1,j} \cdot [a_{ps}]$ |
| Prothrombin | [PT] | $-\varepsilon[PT]\big(\phi_{at}([AP] + [AP_d]) + \phi_{rt}([RP] + [RP_d])\big)$ |
| Thrombin | [TB] | $-\Gamma \cdot [TB] + [PT]\big(\phi_{at}([AP] + [AP_d]) + \phi_{rt}([RP] + [RP_d])\big)$ |
| ATIII | [AT] | $-\Gamma \cdot \varepsilon[TB]$ |

Adhesion of platelets to surfaces and all the other species reactions at a boundary are modeled by surface-flux boundary conditions, following the approach of Sorensen et al. [13,14]. These are listed in Table 4. The specific values and expressions of all the source terms and their parameters are provided in the Table 6. It should be mentioned that for different specific studies the only parameters that need to be determined are the three material surface-specific coefficients, namely $k_{rpd,b}$, $k_{apd,b}$ and $\tau_{emb,b}$. Here, similar to the reaction terms in the internal domain, a negative flux implies consumption and a positive flux implies generation.

Table 4(a). Boundary conditions.

| Species [$C_i$] | $j_i$ form | Description |
|---|---|---|
| [RP] | $-Sk_{rpdb}[RP]$ | Consumption due to [RP]-surface adhesion; Generation due to shear embolization. |
| [AP] | $-Sk_{apdb}[AP]$ | Consumption due to [AP]-surface adhesion; Generation due to shear embolization. |
| [$a_{pr}$] | $\lambda_j\left(k_{apa}[RP_d] + k_{spa}[RP_d] + \theta Sk_{rpdb}[RP]\right)$ | Generation due to agonists and shear activation of [$RP_d$]; Generation due to surface contact activation of [RP]-surface adhesion. |
| [$a_{ps}$] | $s_{pj}[AP_d]$ | Platelet-synthesized generation due to [$AP_d$] |
| [PT] | $-\varepsilon[PT](\phi_{at}[AP_d] + \phi_{rt}[RP_d])$ | Consumption due to thrombin, [TB], generation. |
| [TB] | $[PT](\phi_{at}[AP_d] + \phi_{rt}[RP_d])$ | Generation from prothrombin [PT] due to deposited platelets. |
| [AT] | 0.0 | No reaction flux. |

Table 4(b) Species boundary conditions. (See Table 6 definitions of each of the terms.)

| Species [$C_i$] | $j_i$ form | Description |
|---|---|---|
| [$RP_d$] | $\int_0^t (1-\theta)Sk_{rpdb}[RP] - k_{apa}[RP_d] - k_{spa}[RP_d] - f_{embb}[RP_d]\,dt$ | Generation due to [RP]-surface adhesion; Consumption due to agonists and shear activation; Consumption due to shear embolization. |
| [$AP_d$] | $\int_0^t Sk_{apdb}[AP] + \theta Sk_{rpdb}[RP] + k_{apa}[RP_d] + k_{spa}[RP_d] - (f_{embb} + f_{stb})[AP_d]\,dt$ | Generation due to [AP]-Surface adhesion; Generation due to surface contact activation of [RP]-Surface adhesion; Generation due to agonists and shear activation of [RP]; Consumption due to shear embolization and stabilization. |
| [$AP_s$] | $\int_0^t f_{stb}[AP_d]\,dt$ | Generation due to stabilization. |

The above equations were numerically simulated using the solvers and the libraries of OpenFOAM, a C++ toolbox[66]. The model was then evaluated for two benchmark problems: (1) growth of mural thrombus in an injured blood vessel and (2) deposition within a rectangular crevice. For each geometry studied, the domain was discretized as hexahedral meshes using ICEM. In each of the cases, mesh-dependence studies were performed to assure insensitivity to the mesh size. Numerical results were visualized with ParaView, a post-processing utility for the solution of continuum mechanics problems. [67]

Table 5. Species units, coefficient of species diffusion and initial condition. $\gamma$ is the local shear rate. For more detail see Sorenson[13,14] and Goodman [21].

| Species | Species units | $D_i (m^2 s^{-1})$ | Initial (inlet) condition in blood |
|---|---|---|---|
| [RP] | PLT m$^{-3}$ | $1.58 \times 10^{-13} + 6.0 \times 10^{-13}\gamma$ | $1.5 \times 10^{14} - 3.0 \times 10^{14}$ (Human) <br> $1.9 \times 10^{14} - 10.0 \times 10^{14}$ (Mouse) |
| [AP] | PLT m$^{-3}$ | $1.58 \times 10^{-13} + 6.0 \times 10^{-13}\gamma$ | $0.01[RP] - 0.05[RP]$ |
| [$a_{pr}$] | nmol m$^{-3}$ | $2.57 \times 10^{-10}$ | 0.0 |
| [$a_{ps}$] | nmol m$^{-3}$ | $2.14 \times 10^{-10}$ | 0.0 |
| [PT] | nmol m$^{-3}$ | $3.32 \times 10^{-11}$ | $1.1 \times 10^6$ |
| [TB] | U m$^{-3}$ | $4.16 \times 10^{-11}$ | 0.0 |
| [AT] | nmol m$^{-3}$ | $3.49 \times 10^{-11}$ | $2.844 \times 10^6$ |
| [RP$_d$] | PLT m$^{-3}$ | N/A | 0.0 |
| [AP$_d$] | PLT m$^{-3}$ | N/A | 0.0 |
| [AP$_s$] | PLT m$^{-3}$ | N/A | 0.0 |

Table 6. Value or expression and description of reaction terms and parameters in current paper.

| Terms | Value or expression | units | Description |
|---|---|---|---|
| $k_{apa}$ | $\begin{cases} 0, \Omega < 1.0 \\ \frac{\Omega}{t_{ct}}, \Omega \geq 1.0 \\ \frac{1}{t_{act}}, \frac{\Omega}{t_{ct}} \geq \frac{1}{t_{act}} \end{cases}$ | (s$^{-1}$) | Platelets activation due to agonists; $t_{ct}$ is the characteristic time, which can be used for adjusting the activation rate, and here we choose $t_{ct} = 1s$ as provided Sorensen [13,14]; $\left(k_{apa} = \frac{1}{t_{act}}, \frac{\Omega}{t_{ct}} \geq \frac{1}{t_{act}}\right)$ implies the reaction cannot be faster than platelets physical activation procedure and 99% platelets will be activated during the activation procedure if the agonists or shear stress is large enough. $t_{act}$ is the characteristic time; $t_{act}$ is suggested to range from 0.1s to 0.5s considering the results by Frojmovic et. al [68] and Richardson [69]. |

| Symbol | Expression/Value | Units | Description |
|---|---|---|---|
| $\Omega$ | $\sum_{j=1}^{n_a} w_j \dfrac{a_j}{a_{j,crit}}$ | (N/A) | $a_j$ refers to the concentration of ADP, TxA$_2$ and Thrombin. Value of $w_j$ and $a_{j,crit}$ see Table 7 |
| $k_{spa}$ | $= \begin{cases} \dfrac{1}{t_{ct,spa}} \\ \dfrac{1}{4.0 \times 10^6 \tau^{-2.3}}, t_{ct,spa} > t_{act} \\ \dfrac{1}{t_{ct,spa}}, t_{ct,spa} < t_{act} \end{cases}$ | (s$^{-1}$) | Platelet activation due to shear stress, $\tau$. Expression $t_{ct,spa} = 4.0 \times 10^6 \tau^{-2.3}$ was provided by Goodman[21] and Hellums[70]. |
| $k_{rpd}$ | $div(k_{pd,f}\vec{n})k_{ra}$ | (s$^{-1}$) | Unactivated platelets - deposited activated platelets ([RP]-[AP$_d$]), deposition rate. $f$ refers to the face of a mesh cell and $\vec{n}$ is the unit normal to the face. Details about how this term is calculated are provided in Supplemental Appendix. |
| $k_{ra}$ | $3.0 \times 10^{-6}$ | (m s$^{-1}$) | Constant related to $k_{rpd}$.[13,14] |
| $k_{apd}$ | $div(k_{pd,f}\vec{n})k_{aa}$ | (s$^{-1}$) | Activated platelets-deposited activated platelets ([AP]-[AP$_d$]) deposition rate. Details about how this term is calculated are provided in Supplemental Appendix. |
| $k_{aa}$ | $3.0 \times 10^{-5}$ | (m s$^{-1}$) | Constant related to $k_{apd}$.[13,14] |
| $f_{emb}$ | $Dia_{PLT} \cdot div(k_{emb,f}\vec{n}) \cdot \left(1 - exp(-0.0095 \dfrac{\tau}{\tau_{emb}})\right)$ | s$^{-1}$ | Platelet embolization due to shear stress; Expression $exp(-0.0095\tau)$ was suggested by Goodman[21]. $Dia_{PLT} = 2.78 \times 10^{-6}m$ is the hydraulic diameter of platelets. |
| $k_{emb,f}$ | $k_{emb,f} = \begin{cases} 1, k_{pd,f} > 0 \\ 0, k_{pd,f} = 0 \end{cases}$ | s$^{-1}$ | |
| $\tau_{emb}$ | 30 | dyne cm$^{-2}$ | Platelet shear embolization related constant.[21] |
| $f_{embb}$ | $\left(1 - exp(-0.0095 \dfrac{\tau}{\tau_{embb}})\right)$ | s$^{-1}$ | Platelet embolization in boundary due to shear stress. |
| $\tau_{embb}$ | To be determined, depends on bio-material. | dyne cm$^{-2}$ | Platelet shear embolization related constant. |
| $PLT_{max}$ | $\dfrac{PLT_{s,max}}{Dia_{PLT}}$ | PLT m$^{-3}$ | The maximum concentration of platelets in space. $PLT_{s,max} = 7 \times 10^{10} PLT m^{-2}$ is the total capacity of the surface for platelets; $Dia_{PLT} = 2.78 \times 10^{-6}m$. |
| $\lambda_j$ | $2.4 \times 10^{-8}$ | nmol PLT$^{-3}$ | The amount of agonist $j$ released per platelet. |
| $\theta$ | 1.0 | (N/A) | Platelets activation by contact. |
| $k_{1,j}$ | $\begin{cases} 0.0161 \text{ for TxA2} \\ 0.0 \text{ for ADP} \end{cases}$ | (s$^{-1}$) | The inhibition rate constant of agonist. |
| $s_{pj}$ | $\begin{cases} 9.5 \times 10^{-12} \text{ for TxA2} \\ 0.0 \text{ for ADP} \end{cases}$ | nmol PLT$^{-3}$ s | The rate constant of synthesis of an agonist. |
| $\varepsilon$ | $9.11 \times 10^{-3}$ | nmol U$^{-1}$ | Unit conversion. From NIH units to SI units. |
| $\phi_{at}$ | $3.69 \times 10^{-15}$ | m$^3$ nmol$^{-1}$ PLT$^{-1}$ U s$^{-1}$ | Thrombin generation rate on the surface of activated platelets |
| $\phi_{rt}$ | $6.5 \times 10^{-16}$ | m³ nmol⁻¹ PLT⁻¹ U s⁻¹ | Thrombin generation rate on the surface of unactivated platelets |
| $\Gamma$ | $\dfrac{k_{1,T}[H][AT]}{\alpha K_{AT}K_T + \alpha K_{AT}\varepsilon[TB] + [AT]\varepsilon[TB]}$ | (s⁻¹) | Griffith's template model for the kinetics of the heparin-catalyzed inactivation of thrombin by ATIII. |
| $k_{1,T}$ | 13.333 | (s⁻¹) | A first-order rate constant. |
| $[H]$ | $0.1 \times 10^6$ | nmol m⁻³ | Heparin concentration, assuming specific activity of 300 U mg⁻¹ and molecular weight of 16 kDa.[14,71] |
| $\alpha$ | 1.0 | (N/A) | A factor to simulate a change in affinity of heparin for ATIII when it is bound to thrombin or for thrombin when it is bound to ATIII. |
| $K_{AT}$ | $0.1 \times 10^6$ | nmol m⁻³ | The dissociation constant for heparin/ATIII. |
| $K_T$ | $3.50 \times 10^4$ | nmol m⁻³ | The dissociation constant for heparin/thrombin. |
| $S$ | $1 - \dfrac{[RP_{db}] + [AP_{db}] + [AP_{sb}]}{PLT_{s,max}}$ | (N/A) | Percentage of the wall (boundary) not been occupied by deposited platelets. |
| $k_{rpdb}$ | To be determined, depends on bio-material. | (m s⁻¹) | Unactivated platelet-boundary(wall) deposition rate. |
| $k_{apdb}$ | To be determined, depends on bio-material. | (m s⁻¹) | Activated platelet-boundary(wall) deposition rate. |
| $f_{stb}$ | 0.0 | (s⁻¹) | Deposited activated platelet stabilization rate. |
| $\phi$ | $\dfrac{RP_d + AP_d + AP_s}{PLT_{max}}$ | (N/A) | Volume fraction of deposited platelets |

Table 7. Threshold concentration of that agonist for platelet activation and agonist-specific weight[13,14,21]. It should be noticed that the threshold of the agonist may vary for some certain according to different experimental studies. For ADP see [68,72–75]; for Thrombin see [76–78]; for TxA₂ see [79–81].

| Species | $a_{j,crit}$ | $w_j$ |
|---|---|---|
| ADP($[a_{pr}]$) | $1.00 \times 10^6$ nmol m⁻³ | 1 |
| TxA₂($[a_{ps}]$) | $0.20 \times 10^6$ nmol m⁻³ | 3.3 |
| Thrombin($[TB]$) | $0.1 \times 10^6$ U m⁻³ | 30 |

# Acknowledgement

This research was supported by NIH grant 1 R01 HL089456.

# Author Contributions

W.-T.W. developed the mathematical model and CFD code, performed the simulations and contributed to the preparation of the manuscript. M.A.J. provided the experimental results shown in Figure 9 and Figure 10. W.R.W. supervised these experimental studies. N.A. and M.M. provided supervision of W.-T.W.. J.F.A. contributed to the writing and editing of the manuscript. All the authors contributed to the editing of the manuscript.

## Additional Information

**Competing financial interests**: The authors declare no competing financial interests.

## Appendix 1. Governing questions of chemical/biological species

Provided below are the full set of governing equations in expanded form.

**Unactivated resting PLTs in flow ([RP])**

$$\frac{\partial [RP]}{\partial t} + div(\boldsymbol{v}_f \cdot [RP]) = div(D_P \cdot \nabla[RP]) - k_{apa}[RP] - k_{spa}[RP] - k_{rpd}[RP] \qquad (7)$$

where $\boldsymbol{v}_f$ is the velocity of the fluid (blood), $D_P$ is the diffusivity of the platelets, $k_{apa}$ and $k_{spa}$ are the platelets activation rate due to agonists and shear stress, and $k_{rpd}$ is the deposition rate between [RP] and deposited platelets. The detail definition of the $k_{rpd}$ in mathematical and numerical see Appendix 2.

**Activated platelets in flow [AP]**

$$\frac{\partial [AP]}{\partial t} + div(\boldsymbol{v}_f \cdot [AP]) = div(D_P \cdot \nabla[AP]) + k_{apa}[RP] + k_{spa}[RP] - k_{apd}[AP] \qquad (8)$$

where $k_{apd}$ is the deposition rate between [AP] and deposited platelets. The detail definition of the $k_{apd}$ in mathematical and numerical see Appendix 2.

**Deposited resting platelets [RP$_d$]**

$$\frac{\partial [RP_d]}{\partial t} = (1-\theta)k_{rpd}[RP] - k_{apa}[RP_d] - k_{spa}[RP_d] - f_{emb}[RP_d] \qquad (9)$$

where $(1-\theta)k_{rpd}[RP]$ is the percentage of [RP] activated by contact with biomaterial surfaces and formatted thrombus, $f_{emb}$ is the cleaning rate of deposited platelets due to shear stress.

**Deposited activated platelets [AP$_d$]**

$$\frac{\partial [AP_d]}{\partial t} = \theta k_{rpd}[RP] + k_{apd}[AP] + k_{apa}[RP_d] + k_{spa}[RP_d] - (f_{emb} + f_{stb})[AP_d] \qquad (10)$$

where $f_{stb}$ is the stabilization rate parameter of [AP$_d$], which convert deposited thrombus to be stabilized thrombus which is harder to be cleaned by shear stress.

**Deposited and stabilized platelets [AP$_s$]**

$$\frac{\partial [AP_s]}{\partial t} = f_{stb}[AP_d] \qquad (11)$$

**PLT-released agonists (ADP) [a$_{pr}$]**

$$\frac{\partial [a_{pr}]}{\partial t} + div(\boldsymbol{v}_f \cdot [a_{pr}]) = div(D_{apr} \cdot \nabla[a_{pr}]) \\ + \lambda_j(k_{apa}[RP] + k_{spa}[RP] + k_{apa}[RP_d] + k_{spa}[RP_d] + \theta k_{rpd}[RP]) - k_{1,j}[a_{pr}] \tag{12}$$

where $\lambda_j$ is the amount of agonist $j$ released per platelet and $k_{1,j}$ is the inhibition rate constant of agonist $j$; and $j$ represents ADP for current specie. Therefore $\lambda_j k_{apa}[RP]$ and $\lambda_j k_{apa}[RP_d]$ are the ADP releasing attributed to the activation of unactivated platelets due to agonist, $\lambda_j k_{spa}[RP]$ and $\lambda_j k_{spa}[RP_d]$ are the ADP releasing attributed to the activation of unactivated platelets due to shear stress and $\lambda_j \theta k_{rpd}[RP]$ represents the ADP releasing attributed to the platelets activation due to contact. $k_{1,j}[a_{pr}]$ is the inhibition rate of ADP.

**PLT-synthesized agonists (TxA$_2$) $[a_{ps}]$**

$$\frac{\partial [a_{ps}]}{\partial t} + div(\boldsymbol{v}_f \cdot [a_{ps}]) = div(D_{aps} \cdot \nabla[a_{ps}]) + s_{pj}([AP] + [AP_d]) - k_{1,j} \cdot [a_{ps}] \tag{13}$$

where $s_{pj}$ is the rate constant of synthesis of the agonist $j$; and $j$ represents TxA$_2$ for current specie. Therefore $s_{pj}[AP]$ and $s_{pj}[AP_d]$ represent the rate of synthesis of TxA$_2$ due to activated platelets in flow and deposited activated platelets. $k_{1,j} \cdot [a_{ps}]$ is the inhibition rate of TxA$_2$.

**Prothrombin [PT]**

$$\frac{\partial [PT]}{\partial t} + div(\boldsymbol{v}_f \cdot [PT]) = div(D_{PT} \cdot \nabla[PT]) - \varepsilon[PT](\phi_{at}([AP] + [AP_d]) + \phi_{rt}([RP] + [RP_d])) \tag{14}$$

where $\varepsilon$ is the unit conversion which is from NIH units to SI units, $\phi_{at}$ and $\phi_{rt}$ are the thrombin generation rate constant on the surface of activated platelets and unactivated resting platelets. Therefore $\phi_{at}[AP]$ and $\phi_{rt}[RP]$ are the thrombin generation rate due to activated and resting platelets in flow, while $\phi_{at}[AP_d]$ and $\phi_{rt}[RP_d]$ are the thrombin generation rate due to deposited activated and unactivated platelets.

**Thrombin [TB]**

$$\frac{\partial [TB]}{\partial t} + div(\boldsymbol{v}_f \cdot [TB]) = div(D_{TB} \cdot \nabla[TB]) + [PT](\phi_{at}([AP] + [AP_d]) + \phi_{rt}([RP] + [RP_d])) - \Gamma \cdot [TB] \tag{15}$$

where $\Gamma$ is the Griffith's template model for the kinetics of the heparin-catalyzed inactivation of thrombin by ATIII. Therefore $\Gamma \cdot [TB]$ is the inactivation rate of thrombin by ATIII.

**ATIII [AT]**

$$\frac{\partial[AT]}{\partial t} + div(\boldsymbol{v}_f \cdot [AT]) = div(D_{AT} \cdot \nabla[AT]) - \Gamma \cdot \varepsilon[TB] \tag{16}$$

where $\Gamma \cdot \varepsilon[TB]$ is the consumption rate of ATIII due to inactivation of thrombin.

## Appendix 2. Mathematical and Numerical Considerations for Deposition Rates k$_{rpd}$ and k$_{apd}$.

Provided below are mathematical and numerical considerations about the two terms in equations (1) – (4) above, that represent the rate of deposition to the surface of the thrombus by unactivated resting platelets ($k_{rpd}$) and activated platelets ($k_{apd}$) in the free stream. Referring to the finite volume schematic depicted in Figure S1, the $k_{rpd}$ of the finite-volume-cell 5 is calculated as

$$k_{rpd} = div(k_{pd,f}\vec{\boldsymbol{n}})k_{ra} \tag{17}$$

where $k_{ra}$ is a constant, $\vec{\boldsymbol{n}}$ is the normal to the finite-volume-face, and $f$ refers to the faces shared by adjacent cells 2, 4, 6 and 8. Assuming that thrombus tends to grow layer by layer, when the volume fraction of the deposited platelets of a finite-volume cell $\phi$ is greater than a critical value $\phi_c$ (for example, 0.74, which is the maximum packing fraction for spheres), this cell is able to influence the neighbor cells. Therefore

$$k_{pd,f} = \begin{cases} \frac{[AP_d]_f}{PLT_{max}}, \phi > \phi_c \\ 0, \phi < \phi_c \end{cases} \tag{18}$$

where $\frac{[AP_d]_f}{PLT_{max}}$ represents the percentage of the area occupied by deposited activated platelets at that mesh face. This rule does not apply to the boundary faces.

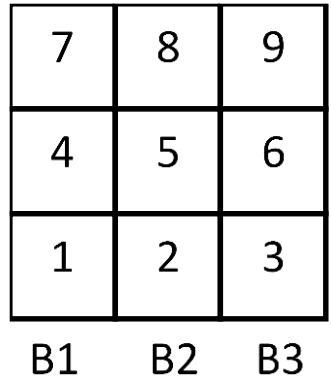

Figure S14 Schematic of cells and faces of a mesh.